\documentclass[%
 reprint,
superscriptaddress,
 amsmath,amssymb,
 aps,
]{revtex4-1}

\usepackage{graphicx}
\usepackage{dcolumn}
\usepackage{bm}
\usepackage{xcolor}
\usepackage{url}
\usepackage{graphicx}
\usepackage{dcolumn}
\usepackage{bm}
\usepackage{amsmath}
\usepackage{epsfig}

\begin{document}

\preprint{APS/123-QED}
\title{Exact solution to the random sequential dynamics of a message passing algorithm} 
\author{Burak \c{C}akmak}
\affiliation{Artificial Intelligence Group, Technische Universit\"at Berlin, Germany}
\author{Manfred Opper} 
\affiliation{Artificial Intelligence Group, Technische Universit\"at Berlin, Germany}%
\affiliation{Centre for Systems Modelling and Quantitative Biomedicine, University of Birmingham, United Kingdom}%
\date{\today}
\def\mathlette#1#2{{\mathchoice{\mbox{#1$\displaystyle #2$}}%
		{\mbox{#1$\textstyle #2$}}%
		{\mbox{#1$\scriptstyle #2$}}%
		{\mbox{#1$\scriptscriptstyle #2$}}}}
\newcommand{\matr}[1]{\mathlette{\boldmath}{#1}}
\newcommand{\RR}{\mathbb{R}}
\newcommand{\CC}{\mathbb{C}}
\newcommand{\NN}{\mathbb{N}}
\newcommand{\ZZ}{\mathbb{Z}}
\newcommand{\bfl}[1]{{\color{blue}#1}}

\newtheorem{assumption}{Assumption}
\newtheorem{theorem}{Theorem}
\newtheorem{remark}{Remark}
\newtheorem{derivation}{Derivation}
\newtheorem{lemma}{Lemma}
\newtheorem{definition}{Definition}
\newcommand{\new}{\color{red}}

%
%
%
%
%

\begin{abstract}
We analyze the random sequential dynamics of a message passing algorithm for Ising models with random interactions in the large system limit. We derive exact results for the two-time correlation functions and the speed of convergence. The {\em de Almedia-Thouless} stability criterion of the static problem is found to be necessary and sufficient for the global convergence of the random sequential dynamics.
\end{abstract}

\maketitle
Probabilistic inference is a key problem in statistics, signal processing and machine learning. 
To make predictions on unobserved random quantities given observed data, averages over conditional distributions have to be computed. For high-dimensional inference problems, the resulting sums or integrals can usually not be performed exactly. To overcome 
this problem, efficient approximate inference algorithms, known as message passing, have been developed \cite{mezard2009information}. Prominent examples are belief propagation and expectation propagation \cite{pearl2014probabilistic,yedidia2005constructing,Minka1,OW5}.
More recently, the so-called {\em approximate message passing} (AMP) algorithms 
designed for probabilistic models on densely connected networks, have been applied
to various inference problems, e.g. Refs. \cite{Kabashima,Donoha, Rangan,krzakala2012statistical,gabrie2015training,Samp,fletcher2018inference,CakmakOpper19,ccakmak2020analysis,cakmak_2020}.  Methods of statistical physics have played an important role in the design and the theoretical analysis of such algorithms, e.g., fixed points of AMP--style algorithms were shown to coincide with the solutions of 
the advanced  {\em Thouless-Anderson-Palmer} (TAP) mean-field equations developed in the statistical physics of disordered systems \cite{Mezard}. This shows that under certain 
statistical assumptions on network couplings, message passing algorithms
can achieve exact predictions in the thermodynamic limit of large systems.
Using techniques of information theory and statistical physics,  exact solutions for the  {\em dynamics} of the AMP-style message passing iterations have also become possible \cite{Bolthausen,Bayati,Opper16,Burak17,rangan2019vector,takeuchi,fan2020approximate}. So far, the theoretical analysis has concentrated on the simplest type of dynamics, the {\em parallel update} of all dynamical variables or nodes in the algorithm. 

In practical applications, however, a sequential update of individual or groups of variables may often be preferable to obtain a more stable behavior. In fact, Tom Minka's expectation propagation (EP) algorithm \cite{Minka1}, which is one of the motivations behind the vector-AMP (VAMP) approach \cite{Ma,rangan2019vector,takeuchi}, is formulated in terms of sequential iterations. Parallel versions of EP often require extra damping procedures (see, e.g., Ref. \cite{vehtari2020expectation}) to achieve convergence. The second advantage of sequential algorithms over parallel ones might be a reduced computational complexity. In the case of the Ising model, for example, sequential updating of individual variables reduces the need for matrix-vector multiplication in the parallel updates to vector-vector multiplication at each iteration step.

In this paper, we obtain an exact large-system analysis of the dynamics
for an AMP-style message passing algorithm with random sequential updates. We show that the effective dynamics of a single node is described by a simple stochastic equation driven by a Gaussian process. We derive explicit analytical conditions for global convergence and compute the convergence time.
This result is nontrivial, because previous studies of other random sequential learning algorithms, e.g., Refs. \cite{sollich1997line,mignacco2020dynamical} have shown that the effective
single node dynamics can be more complex due to the occurrence of memory terms. 

For simplicity, we will focus on a class of toy problems for inference which is given 
by the prediction of magnetizations $\matr {m}=\mathbb E[\matr s]$ for Ising models with pairwise interactions between spin variables $\matr s=(s_1,\ldots,s_N)^\top\in\{-1,1\}^{N}$. 
Generalizations to teacher-student scenarios for other inference problems such as generalized linear models will be discussed in a future publication. For applications of Ising models to real data, see, e.g., Ref. \cite{nguyen2017inverse}. 
The Ising model is defined by the Boltzmann distribution:
 \begin{equation}
p(\matr s\vert \matr J,h)\doteq \frac{1}{Z}\exp\left(\sum_{i,j\leq N}J_{ij}s_is_j+ h\sum_{i\leq N}s_i\right)\label{Gibbs}.
\end{equation}
To discuss a typical inference task, we assume that the coupling matrix $\matr J=\matr J^{\top}$ is drawn from an arbitrary rotation-invariant random matrix ensemble. This means that $\matr J$ and $\matr O\matr J\matr O^\top$ have the same probability distribution for any orthogonal matrix $\matr O$ independent of~$\matr J$. This leaves the freedom to specify the spectrum
of the matrix~$\matr J$. A special case of this ensemble is given by independent 
zero-mean Gaussian couplings, known as the {\em Sherrington-Kirkpatrick} (SK) model \cite{SK}. 
In general, however, matrix elements are statistically dependent for a rotation invariant ensemble. 

We consider approximations of the magnetizations $\matr m$ which are given by the so-called TAP mean-field equations. For invariant random coupling matrices, these 
are  given by \cite{Parisi,Adatap}
\begin{subequations}
\label{tap} 
	\begin{align}
		\matr m&={\rm Th}(\matr\gamma)\label{tap1},\\
		\matr \gamma&=\matr J\matr m-{\rm R}(\chi)\matr m.
		\end{align} 
\end{subequations}
Here, for short, we have defined the non-linear function ${\rm Th}(x)=\tanh(h+x)$ and 
$\chi \doteq \mathbb E[{\rm Th}'(\sigma_\gamma u)]$, where $u$ is a zero-mean normal Gaussian random variable and $\sigma_\gamma^2\doteq (1-\chi){\rm R}'(\chi)$. The function ${\rm R}$ stands for the R-transform \cite{mingo2017free} of the limiting spectral distribution of $\matr J$ defined as
$
{\rm R}(\omega)\doteq {\rm G}^{-1}(\omega)-1/\omega,  
$
where ${\rm G}^{-1}$ is the functional inverse of the Green's function 
$
{\rm G}(z)\doteq \lim_{N\to\infty}\mathbb E[(z{\bf I}-\matr J)^{-1}_{ii}]$. To ensure that the Green's function has a unique
inverse, we assume $\chi<\lim_{z\to \lambda_{+}}{\rm G}(z)$, where $\lambda_{+}$ stands for the supremum  of the support of the limiting spectral distribution of $\matr J$. To define an AMP-style algorithm for solving the TAP equations, we first transform \eqref{tap}
into an equivalent, canonical form 
 \begin{align}
\matr A f(\matr \gamma) = \matr \gamma.
\label{tap_mf1}
\end{align}
The function $f$ is applied component wise to the vector 
$\matr  \gamma$ and $\matr A$ is a $N\times N$ matrix. The two conditions on this transformation
which are essential for the further analysis are that
\begin{equation}
\mathbb E[{f'}(\sigma_\gamma u)] = 0, \qquad
\lim_{N\to \infty} \mathbb E[(\matr A)_{ii}] =0 \label{require}
\end{equation}
together with the fact that $\matr A$ is a random matrix with rotationally
invariant distribution. For the Ising problem, 
this is achieved by setting $\matr m = \chi({\matr \gamma} + f(\matr \gamma))$ and by using the definitions 
\begin{subequations}
\label{equivtap}
	\begin{align}
	f(x)&=\frac{1}{\chi}{\rm Th}(x)-x \\
	\matr A&=\frac{1}{\chi}({\rm G}^{-1}(\chi){\bf I}-\matr J)^{-1}-{\bf I}.
\end{align}
\end{subequations}
While \eqref{equivtap} are specific to the Ising problem, similar transformations are possible 
for other inference problems. We define an AMP-style iterative algorithm for solving (\ref{tap_mf1}) in discrete time 
 $k=1,2,\ldots$ by
\begin{subequations}
	\label{vamp}
	\begin{align}
		\matr \phi^{(k)}&=\matr Af(\matr \gamma^{(k-1)})\label{phil}\\
		\matr \gamma^{(k)}&=\matr \gamma^{(k-1)}+\matr P^{(k)}[\matr \phi^{(k)}-\matr \gamma^{(k-1)}]\label{vamp1}.
	\end{align}
\end{subequations}
The initialization is given by
$
\matr \gamma^{(0)}=\sigma_\gamma \matr u
$
where $\matr u$ is a vector with independent zero-mean normal 
random variables;
(\ref{vamp1}) is  a generalization of the
parallel iterative algorithm given in Ref. \cite{CakmakOpper19} which 
was motivated by the VAMP algorithms of Refs. \cite{Ma,rangan2019vector,takeuchi}. 
The parallel dynamics of Ref. \cite{CakmakOpper19} is obtained when the diagonal matrix $\matr P^{(k)}$ is equal to the unit matrix. By introducing {\em binary} diagonal entries
$p_i^{(k)}\doteq P^{(k)}_{ii}\in\{0,1\}$, we obtain random {\em sequential} updates 
of nodes. The random decision variables decide 
if node $i$ is updated ($p_i^{(k)} =1$) at time $k$ or not  ($p_i^{(k)} =0$).
We assume that the $p_i^{(k)}$ are independent for all $i,k$  and that
$ {\rm  Pr}(p_i^{(k)}=1) = \eta$. The case
$\eta = 1/N$ corresponds to an update of only a single node on average.
  
We will next derive the statistical properties of the dynamics \eqref{vamp}
in the thermodynamic limit of large $N$ while keeping 
$\eta$ fixed. We will later also discuss the limit $\eta\to 0$ to simulate 
the behavior for $\eta =1/N$.

Our goal is to show that for $N\to\infty$, the 
sequence $\{\phi_i^{(k)}\}_{k=1}^K$ over $K$ time steps 
for an arbitrary component $i$ converges to a zero mean Gaussian process. 
We will build on results of
Ref. \cite{Opper16} which are based on the {\em dynamical functional theory}
of statistical physics. 
This path integral method allows for an explicit averaging over the 
randomness of the matrix $\matr A$ and leads to a decoupling of the degrees of freedom.
Using the second condition \eqref{require} for the random matrix $\matr A$ it was shown in Ref. \cite{Opper16} that 
$\{\phi^{(k)}\}_{k=1}^K$ (suppressing the component index $i$ for convenience)
can be transformed into a Gaussian random sequence by appropriate {\em subtractions}. The 
subtractions define an auxiliary dynamical system which is obtained 
by replacing the variable $\matr \phi^{(k)}$ in \eqref{phil}  by 
\begin{equation}
\matr \phi_{\rm aux}^{(k)}= \matr A f(\matr \gamma^{(k-1)})-\sum_{l<k}\hat{\mathcal G}^{(k,l)}f(\matr \gamma^{(l-1)})
\label{auxdyn}
\end{equation}
for $k=1,2,\ldots K$. Under the new dynamics,  $\{\phi_{\rm aux}^{(k)}\}_{k=1}^K$
can be shown to be a Gaussian process. 
The memory terms in (\ref{auxdyn}) are defined as follows:
$\hat{\mathcal G}^{(k,l)}$ denotes the $(k,l)$th indexed entries of the $K\times K$ matrix $\hat{\mathcal G}$ which is defined in terms of the R-transform 
and its power series expansion as
\begin{equation}
\hat{\mathcal G}={\rm R}_\matr A(\mathcal G)=\sum_{n=1}^{\infty}c_{\matr A,n}\mathcal G^{n-1}\label{reseq}.
\end{equation}
Finally, the entries of the {\em response matrix} $\mathcal G$ are given by
\begin{equation}
\mathcal G^{(k,k')}\doteq \lim_{N\to \infty}\mathbb E\left[\frac{\partial f(\gamma^{(k-1)})}{\partial \phi_{\rm aux}^{(k')}}\right]
\end{equation}
again suppressing  the component index $i$ for convenience, i.e., $\gamma^{(k)} = \gamma_i^{(k)}$ of $\matr  \gamma^{(k)}$ and  $\phi_{\rm aux}^{(k)} = \phi_{\rm aux,i}^{(k)}$ of
	$\matr  \phi_{\rm aux}^{(k)}$.
We will show next,  that $\mathcal G=\matr 0$. From this we also obtain
$\mathcal {\hat G}= \matr 0$. This will prove that 
$\matr \phi^{(k)} = \matr \phi_{\rm aux}^{(k)}$ 
and \eqref{auxdyn} reduces to \eqref{phil}. 
By construction we have 
\begin{equation}
\frac{\partial{\gamma}^{(k-1)}}{\partial {\phi}_{\rm aux}^{(k')}}=\underbrace{p^{(k')}\prod_{l=k'+1}^{k-1}(1-p^{(l)})}_{\doteq p^{(k,k')}}\quad~ k'<k.
\end{equation}
Hence, the response terms read 
\begin{align}
&\mathcal G^{(k,k')}=\lim_{N\to \infty}\mathbb E\left[f'(\gamma^{(k-1)})p^{(k,k')}\right]\nonumber \\
&={\rm Pr}( p^{(k,k')}=1)\lim_{N\to \infty}\mathbb E\left[f'(\gamma^{(k-1)})p^{(k,k')}\vert p^{(k,k')}=1\right]\nonumber \\	&=\eta(1-\eta)^{k-1-k'}\lim_{N\to \infty}\mathbb E\left[f'(\phi_{\rm aux}^{(k')})\right].
\end{align} We will only sketch the the final step of the proof. It is based on
a careful analysis of the two-time covariance function of the Gaussian process 
$\matr \phi_{\rm aux}^{(k)}$ (see Ref. \cite{Opper16}),
\begin{equation}
\mathcal C_{\phi_{\rm aux}}=\sum_{n=2}^{\infty}c_{\matr A,n}\sum_{k=0}^{n-2} \mathcal {G}^k\mathcal {C}_f(\mathcal {G}^\top)^{n-2-k},\label{covphi}
\end{equation}
where 
\begin{equation}
\mathcal  C_f^{(k,k')}= \lim_{N\to \infty}\mathbb E[f(\gamma^{(k-1)})f(\gamma^{(k'-1)})].
\end{equation}
One can show 
 by induction (starting with the initialization $\mathcal C_{\gamma}^{(0,0)} =  \sigma_{\gamma}^2$) that
 the variances of $\matr \phi_{\rm aux}^{(k)}$ are constant in time, i.e.
\begin{align}
\mathcal C_{\phi_{aux}}^{(k,k)} = \sigma_\gamma^2\label{equalvariance}.
\end{align}
Hence, using the condition \eqref{require}, we obtain
\begin{equation}
\lim_{N\to \infty}\mathbb E\left[f'(\phi_{\rm aux}^{(k)})\right]=0 \label{dfc}
\end{equation}
which establishes the vanishing of memory terms and Gaussianity of 
$\{\phi^{(k)}\}_{k=1}^K$.

Hence, as the main result of our paper, we have shown that the effective dynamics of a 
single node of the algorithm is given by the stochastic dynamical equation
 \begin{align}
\gamma^{(k+1)}= \gamma^{(k)}+
p^{(k+1)}[\phi^{(k+1)} - \gamma^{(k)}]
\label{effective_dyn}
\end{align}
where the temporal sequence $\{\phi^{(k)}\}_k$ is a Gaussian random process.
The vanishing of the response terms ${\mathcal G}$ also leads to a simplification of the two-time covariances
\begin{align}
	\mathcal C_{\phi}^{(k,k')}&= 
c_{\matr A,2} \mathcal C_{f}^{(k,k')} =  \mathcal C_{f}^{(k,k')}
	\lim_{N\to\infty} \mathbb E[(\matr A^2)_{ii}]\;,\label{twotimephi}
\end{align}
where the latter equality follows from properties of the R--transform.
This result together with the fact that the binary decision variables $p^{(k)}$ are independent for different times and also independent of the Gaussian process, specifies the statistics of the single node trajectories 
$\{\gamma^{(k)}\}_{k=1}^K$ completely.  Although the joint distribution of the random variables  $\gamma^{(k)}$ and $\gamma^{(k')}$ (for any $k\neq k'$) is non--Gaussian, the {\em  linearity}  of the dynamics (\ref{effective_dyn}) 
allows for a simple recursive computation of 
moments at different times in terms of the moments of the driving Gaussian
variables. 
For $k\neq k'$, one obtains the recursions 
\begin{align}
&\mathcal C_{\phi}^{(k,k')}=(1-\eta)^2\mathcal C_{\phi}^{(k-1,k'-1)}+\eta^2\left[\mathcal C_{\tilde \phi}^{(k,k')}\right.\nonumber\\
&\left.+\sum_{l'=1}^{k'-1}(1-\eta)^{k'-l'}\mathcal C_{\tilde \phi}^{(k,l')}+\sum_{l=1}^{k-1}(1-\eta)^{k-l}\mathcal C_{\tilde \phi}^{(k',l)}\right]\;, \label{twotimegamma}
\end{align}
where we have introduced the two-time expectations
\begin{align}
	\mathcal C_{\tilde\phi}^{(k,k')}&=c_{\matr A,2}\mathbb E[f(\phi^{(k-1)})f(\phi^{(k'-1)})]\quad  k,k'>1,\\
	\mathcal C_{\tilde \phi}^{(k,1)}&=\frac{c_{\matr A,2}(\mathbb E[f(\sigma_\gamma u)])^2} \eta\;,~~\qquad~ \qquad k\neq 1.
\end{align}
We obtain similar recursions for  the two-time covariances 
\begin{align}
&\mathcal C_\gamma^{(k,k')}=(1-\eta)^2\mathcal C_\gamma^{(k-1,k'-1)}+\eta^2\left[\mathcal C_\phi^{(k,k')}\right.\nonumber\\
&\left.+\sum_{l'=1}^{k'-1}(1-\eta)^{k'-l'}\mathcal C_\phi^{(k,l')}+\sum_{l=1}^{k-1}(1-\eta)^{k-l}\mathcal C_\phi^{(k',l)}\right]\;,\label{covgamma}
\end{align}
with $\mathcal C_\gamma^{(k,0\;)}=(1-\eta)^{k}\sigma_\gamma^2$. Moreover, the variances read
\begin{align}
\mathcal C_\gamma^{(k,k)}&=\eta\mathcal C_\phi^{(k,k)}+(1-\eta)\mathcal C_\gamma^{(k-1,k-1)}=\sigma_\gamma^2 \;,\label{var2}
\end{align}
where the latter equality follows from \eqref{equalvariance} by induction.

To analyze the convergence properties of the dynamics \eqref{vamp} we consider
the limit of the two-time covariances, when one time index approaches infinity. Setting
$\mathcal C_{\gamma,\phi}^{(k)} \doteq \lim_{k'\to\infty} \mathcal C_{\gamma,\phi}^{(k,k')}$, 
one can show 
from the recursions \eqref{twotimegamma} and \eqref{covgamma} that
\begin{subequations}
	\label{difkrecurs}
	\begin{align}
		\mathcal C_\gamma^{(k)}&=\eta \mathcal C_\phi^{(k)}+(1-\eta) {\mathcal C}_\gamma^{(k-1)}\label{Cgam}\;,\\
		\mathcal C_{\phi}^{(k)}&=\eta g(\mathcal C_\phi^{(k-1)})+(1-\eta) {\mathcal C}_{\phi}^{(k-1)}\;,\label{Cphi}
	\end{align} 	
\end{subequations}
with the necessary initial values $\mathcal C_\gamma^{(0)}=0$ and $\mathcal C_{\phi}^{(1)}=g(0)$. Here, we have introduced the function
\begin{equation}
g(x)\doteq c_{\matr A,2}\mathbb E[f(\phi_1)f(\phi_2)]\label{gfunc}
\end{equation} 
for $\phi_1$ and $\phi_2$ being jointly Gaussian random variables with covariance $x$ and equal variances $\sigma_\gamma^2$.  This enables us to study the deviation between 
variables at time $k$ and their long-time limits:
\begin{equation}
\Delta_\gamma^{(k)}\doteq \lim_{k'\to \infty} 
\lim_{N\to\infty}\frac{1}{N}\mathbb E[\Vert \matr \gamma^{(k)}-\matr \gamma^{(k')} \Vert^2] = 2\sigma_\gamma^2-2\mathcal C_\gamma^{(k)}.
\end{equation}
One can show that global convergence of the algorithm is achieved under the condition
\begin{equation}
\lim_{k\to \infty}\Delta_\gamma^{(k)}=0 \iff g'(\sigma_\gamma^2)<1\;,
\label{cond2}
\end{equation}
independent of the probability $\eta$ for an update. Following Ref. \cite{CakmakOpper19}, 
where parallel updates ($\matr P^{(k)} ={\bf I}$) were analyzed, we
can show that the condition for convergence \eqref{cond2} coincides with the well-known \emph{de Almedia-Thouless} (AT) stability criterion \cite{AT} of the replica-symmetric solution of Ising models with rotation invariant coupling matrices \cite[Eq. (46)]{Enzo}. 
It is interesting to note that for $g'(\sigma_\gamma^2)\geq 1$ the algorithm fails to converge
although the variance of $\gamma^{(k)}$ remains constant in time.

\begin{figure*}	
	\includegraphics[width=1\textwidth]{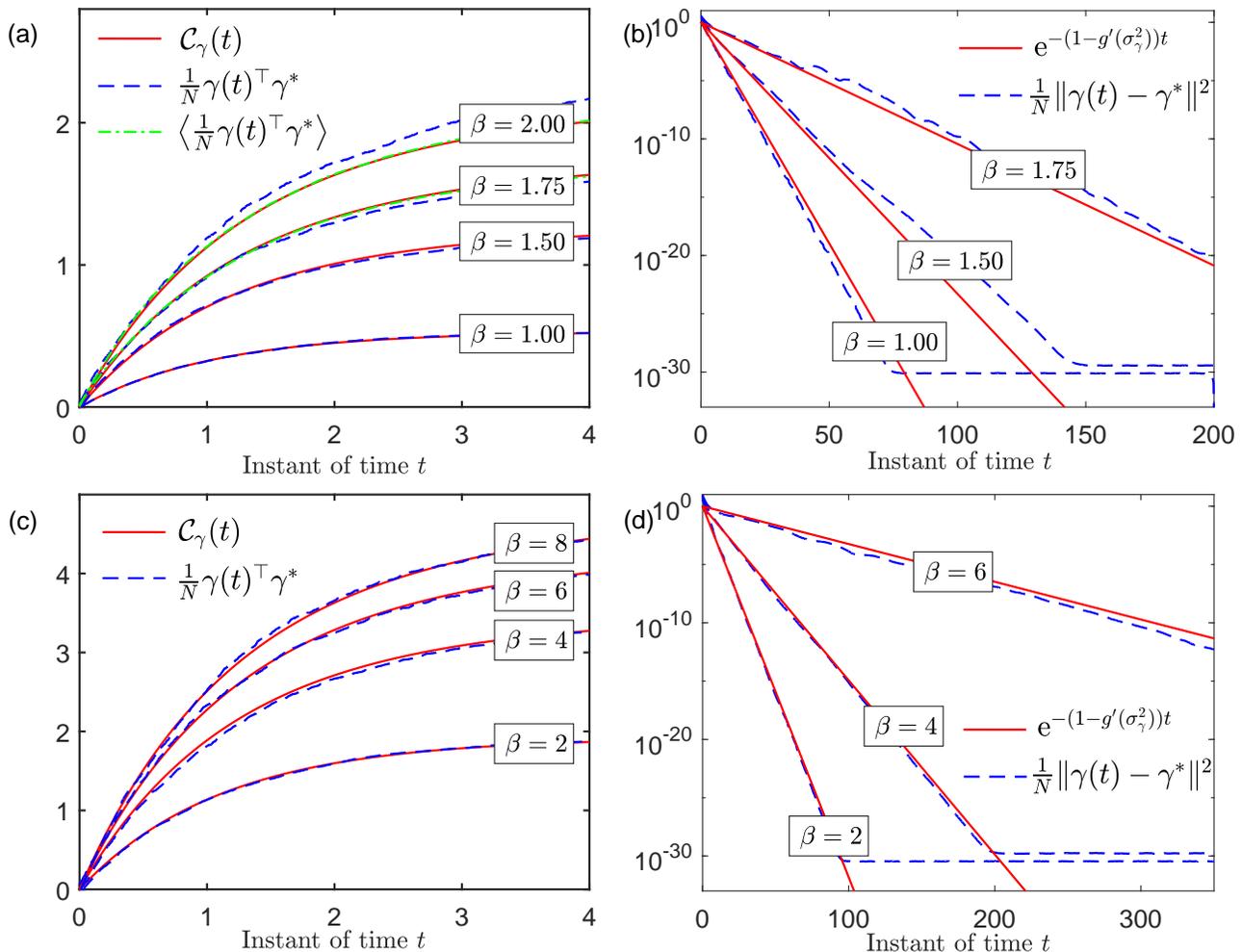}
	\caption{Comparison of theory and simulations for $\eta=1/N$=$10^{-4}$ and $\matr \gamma(t)\doteq \matr \gamma^{(\frac{t}{\eta})}$ for $\frac{t}{\eta}=0,1,2,\ldots$: Fig. (a) and (b) are for the SK model with $h=1$. $\beta=1.91$ gives the AT line of instability ($g'(\sigma_\gamma^2)=1$). Fig. (c) and (d) are for the random-orthogonal-model with $h=2$. 
		$\beta=6.70$ gives the AT line of instability. In the region of stability, $\matr \gamma^*\doteq\matr \gamma(t_*)$ denote stationary vectors for sufficiently large times $t_\star$. Otherwise, we chose $\matr \gamma^*=\matr \gamma (100)$. The empirical averages $\langle \cdot  \rangle$ are computed over ten realizations of the dynamics. Flat lines around $10^{-30}$ are due to the machine precision of the computer which was used.}
	\label{fig1}
\end{figure*}
We will now specialize to the case where only a small number of nodes
is updated. This limit is interesting for practical applications of algorithms.
For simplicity, we consider $\eta = 1/N$ where a single node is updated on average. 
To model such a process within our approach, we take the limit $\eta \to 0$ and introduce 
a re--scaling of time $t = \eta k$ which becomes a continuous variable in the limit.
We write (with a slight abuse of notation) $\gamma(t)$ and $\phi(t)$ instead of $\gamma^{(k)}$
and $\phi^{(k)}$, etc.  The discrete recursions \eqref{difkrecurs} are then replaced by the 
ordinary differential equations 
\begin{subequations}
\label{ODEs}
\begin{align}
\dot{\mathcal C}_\gamma(t)&=\mathcal C_\phi(t)-\mathcal C_\gamma(t)\;,\label{ode1}\\
\dot{\mathcal C}_\phi(t)&=g (\mathcal C_{\phi}(t))-\mathcal C_{\phi}(t)\;,\label{ode2}
\end{align}	
\end{subequations}
where the dots denote derivatives with respect to time $t$. Linearizing the function
$g(x)$ around the fixed point, we obtain
the asymptotic solution 
\begin{equation}
\Delta_\gamma(t)\simeq {\rm e}^{-(1-g'(\sigma_\gamma^2))t} \label{dev}
\end{equation}
for $t\to\infty$ if $g'(\sigma_\gamma^2)< 1$.
This again manifests the AT line of stability \eqref{cond2} as the sufficient and necessary condition for the global convergence. In Fig.~\ref{fig1}, we illustrate the theoretical predictions of the results \eqref{ODEs} and \eqref{dev}. We  consider two random coupling matrix models: the SK-model where the couplings are independent 
Gaussian entries with zero mean and the variances $\mathbb E[J_{ij}^2]=\beta(1+\delta_{ij})/N$; a random-orthogonal-model \cite{Parisi}
for which the eigenvalues of the coupling matrix are binary $\mp \beta$ with the trace-free property ${{\rm tr}}(\matr J)=0$ whenever $N$ is an even number. The simulation results are (mainly) based on single realizations of the dynamics but different realizations are considered for each value of the inverse temperature $\beta$. 
As the model parameters approach (are) to (in) the region of dynamical instability, the discrepancy between the theory and simulations may increase due to the fluctuation of the realizations, e.g., for the SK model, we illustrate the theoretical predication of \eqref{ODEs} through an empirical average over a number of realizations of the dynamics, as well. On the other hand,  for the second model the theoretical results already give excellent agreement with a single realization of the dynamics. This might stem from the fact that the system shows smaller fluctuations as the random matrix has a nonrandom spectral distribution for finite $N$.

We analyzed the dynamics of a message passing algorithm 
for approximate inference with random sequential updates in the thermodynamic limit. By deriving an effective stochastic dynamics for a single node, we were able to
obtain explicit results for the asymptotic convergence. For simplicity,  to 
demonstrate our main ideas, we have restricted our analysis in two ways: We considered an Ising model as a toy inference problem. We also 
specialized to a simplified AMP-style algorithm which starts with the proper initialization to keep the variance of variables constant in time. With a bit more technical effort, both restrictions can be easily lifted.  Our analysis can e.g. be extended to the common teacher-student scenario for generalized linear data models \cite{mccullagh2018generalized}. The inclusion of more adaptive updates used e.g. in VAMP algorithms \cite{Ma,rangan2019vector,takeuchi} is also possible and will be given in a forthcoming publication.

From a theoretical point of view, we expect that most of our analysis can be made mathematically
rigorous using, e.g., the recent approach \cite{fan2020approximate} to justify the subtraction rule \eqref{auxdyn}. There is, however, a subtle point related to the limit $\eta = 1/N$ of 
single node updates which might need further investigation. The dynamical functional approach
used to derive our results is restricted to the limit $N\to\infty$, but with the number of time steps 
$K$ kept finite. For $\eta \propto 1/N$, we also need to increase the number of iterations
$K\propto N$ in order to have nonzero changes in the dynamics. Although
our results are supported very well by simulations, we may try an alternative approach, where
the continuous time limit $\eta\to 0$ in the dynamical functional theory is performed {\em before} the limit $N\to\infty$. The discrete time decision variables would then be replaced by Poisson events.
We leave this calculation to subsequent publication but conjecture that the resulting
ordinary differential equations would agree with \eqref{ODEs}.

This work was supported by the German Research Foundation, Deutsche Forschungsgemeinschaft (DFG), under Grant ``RAMABIM'' with No. OP 45/9-1.

\bibliography{mybib}

\end{document}